\documentclass[usenatbib,longnamesfirst]{mn2e}
\RequirePackage{graphicx}
\begin{document}
\title[A SCUBA survey of L1689]{A SCUBA survey of L1689 -- the dog that 
didn't bark}
\author[Nutter, Ward-Thompson \& Andr\'{e}]
{D. Nutter$^1$\thanks{E-mail: David.Nutter@astro.cf.ac.uk}, 
D. Ward-Thompson$^1$,
P. Andr\'{e}$^2$ \\
$^1$Department of Physics and Astronomy, Cardiff University, 5 The Parade, 
Cardiff, CF24 3YB\\
$^2$CEA, DSM, DAPNIA, Service d'Astrophyique, C.E. Saclay, F-91191 
Gif-sur-Yvette Cedex, France}

\maketitle

\begin{abstract}
We present submillimetre data for the L1689 cloud in the $\rho$~Ophiuchi 
molecular cloud complex. We detect a number of starless and prestellar 
cores and protostellar envelopes. We also detect a number of filaments 
for the first time in the submillimetre continuum that are parallel both 
to each other, and to filaments observed in the neighbouring L1688 cloud. 
These filaments are also seen in the $^{13}$CO observations of L1689. The 
filaments contain all of the star-formation activity in the cloud. L1689 
lies next to the well studied L1688 cloud that contains the $\rho$~Oph-A 
core. L1688 has a much more active star-formation history than L1689 despite 
their apparent similarity in $^{13}$CO data. Hence we label L1689 as the dog that didn't bark. We endeavour to explain this 
apparent anomaly by comparing the total mass of each cloud that is 
currently in the form of dense material such as prestellar cores. We note firstly that L1688 is more massive than L1689, but we also find that when normalised to the total mass of each cloud, the L1689 cloud has a 
much lower percentage of mass in dense cores than L1688.
We attribute this to the hypothesis of \citet{1989ApJ...338..902L} 
that the star formation in the $\rho$~Ophiuchi complex is being affected and 
probably dominated by the external influence of the nearby Upper Scorpius 
OB association and predominantly by $\sigma$~Sco. L1689 is further from 
$\sigma$~Sco and is therefore less active. The influence of $\sigma$~Sco 
appears nonetheless to have created the filaments that we observe in L1689.
\end{abstract}


\section{Introduction}\label{intro}

The earliest stages of low-mass (0.2--2M$_\odot$) star formation are becoming 
reasonably well understood (see e.g. \citealp{2000prpl.conf...59A} for a 
review). The prestellar core stage \citep{1994MNRAS.268..276W} is the phase 
in which a molecular cloud core has become gravitationally bound. Thereafter 
gravitational collapse sets in and a central hydrostatic protostar forms, 
which is known as a Class 0 protostar \citep{1993ApJ...406..122A}. Once half 
of the final mass has accreted onto the central object it is known as a Class 
I protostar \citep{1989ApJ...340..823W}, and it subsequently evolves through 
the Class II \& III young stellar object (YSO) phases 
\citep{1987IAUS..115....1L}. Debate continues over the details of this 
evolutionary process, and different molecular clouds appear to be evolving 
according to a different interplay of physical mechanisms. These include 
magnetic fields, turbulence and feedback from previous episodes of star 
formation. Submillimetre studies of different star-forming regions are 
required to help clarify the evolutionary process and determine the 
effect of these mechanisms.

\begin{figure*}
\includegraphics[angle=270,width=150mm]{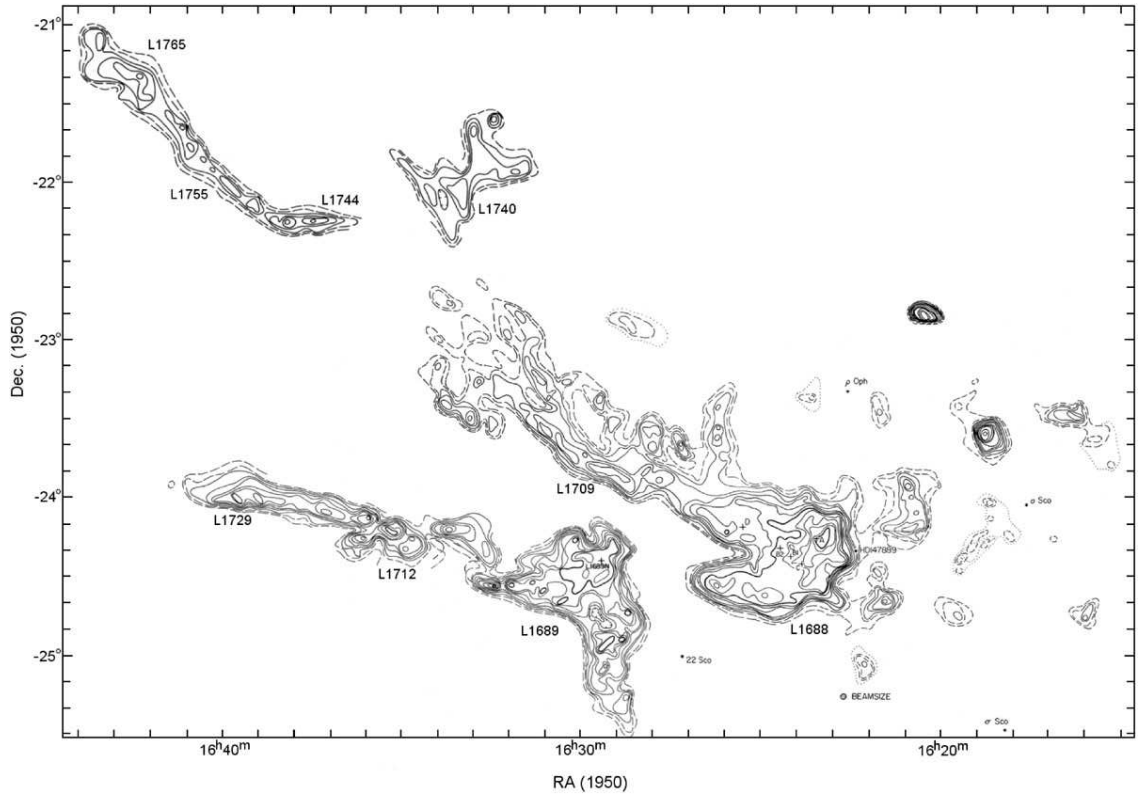}
\caption{$^{13}$CO map of the $\rho$~Ophiuchi complex, 
from \citet{1989ApJ...338..902L}. The contour levels give antenna 
temperatures $T_R^\star(^{13}{\rm CO})$ of 4, 5, 6, 7, 8, 10, 12, 14, 18 
and 20~K. The names of the clouds are marked.}
\label{loren}
\end{figure*}

The L1689 molecular cloud is part of the $\rho$~Ophiuchi molecular cloud 
complex, which is located at a distance of 128 $\pm$ 12 pc from the sun 
\citep{1999A&A...352..574B}. It was first detected by 
\citet{1962ApJS....7....1L} in a large study of the dark nebulae detected 
in the Palomar Observatory Sky Survey plates. The $\rho$~Ophiuchi complex 
was extensively mapped by \citet{1989ApJ...338..902L} using $^{13}$CO at a 
resolution of 2.4 arcmin. The resulting map of the cloud is shown in 
Fig.~\ref{loren}. The data show that the molecular cloud is composed 
of a number of sub-clouds, the most massive of which is L1688. 

\begin{figure*}
\includegraphics[angle=0,width=113mm]{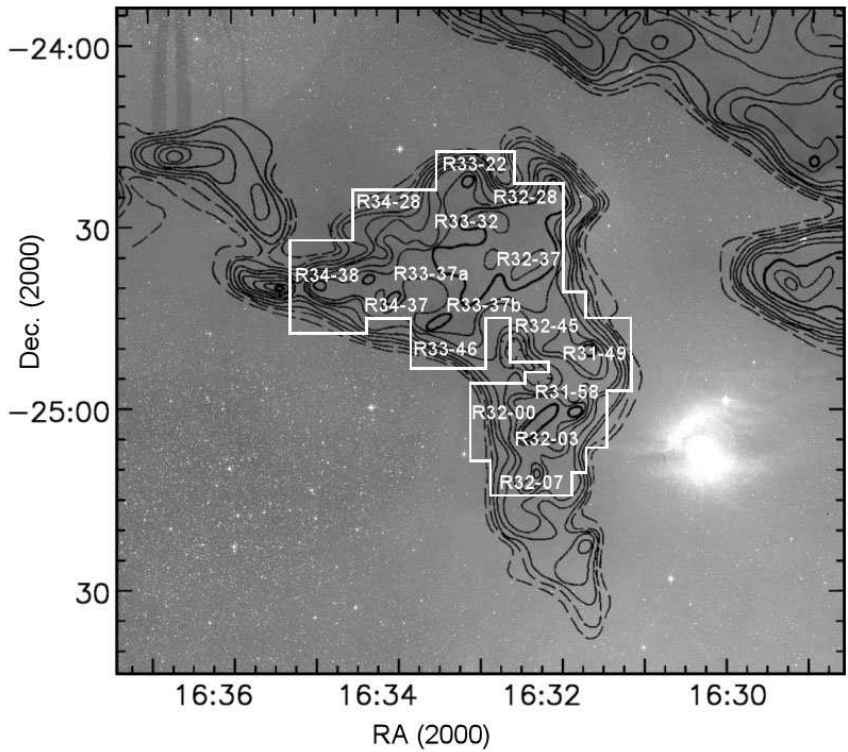}
\caption{A composite map of the L1689 cloud showing the scan-mapped 
fields (white outline and labels). The background 0.5~$\mu$m image is 
taken from the Digital Sky Survey, obtained using the {\em SkyView} 
interface. The contours are $^{13}$CO emission at $T_R^\star(^{13}{\rm CO})$ 
of 4, 5, 6, 7, 8, 10, 12, 14, 18 and 20~K \citep{1989ApJ...338..902L}.}
\label{overlay}
\end{figure*}

\begin{table*}
\begin{center}
\caption{The positions of each of the scan-mapped fields in the L1689 
molecular cloud. The atmospheric conditions during each observation are 
indicated by the 850 $\mu$m zenith optical depth listed in column~5. 
The field names correspond to those shown in Fig.~\ref{overlay}.}
\label{obs_table}
\vspace{0.5cm}
\begin{tabular}{lcccccccc}\hline
	& \multicolumn{2}{c}{Map centre}&		&	\\ 
\cline{2-3}
Field 	& RA		& Dec. 	 	& 		& 	\\
Name	& (2000)	& (2000)	& UT Date	& 
$\tau_{850 \mu m}$\\ \hline
R31-49	&16:31:39.6	&$-$24:49:43	&1999-Mar-09	& 0.34\\
~~~~--	&--		&--		&2000-Apr-09	& 0.26\\
R31-58	&16:31:56.7	&$-$24:58:52	&1999-Mar-09	& 0.38\\
~~~~--	&--		&--		&2000-Apr-10	& 0.22\\ 
R32-45	&16:32:11.5	&$-$24:45:21	&2000-Apr-09	& 0.22\\ 
R32-03	&16:32:14.1	&$-$25:03:10	&1999-Aug-08	& 0.22\\
~~~~--	&--		&--		&1999-Aug-09	& 0.38\\ 
R32-07	&16:32:22.7	&$-$25:07:50	&1999-Apr-08	& 0.22\\
~~~~--	&--		&--		&2000-Apr-10	& 0.18\\ 
R32-37	&16:32:26.1	&$-$24:37:20	&2000-Apr-09	& 0.22\\
R32-28	&16:32:27.6	&$-$24:28:25	&1999-Mar-09	& 0.30\\
R32-00	&16:32:38.8	&$-$25:00:49	&2000-Apr-10	& 0.18\\
R33-22	&16:33:01.9	&$-$24:22:17	&2000-Apr-12	& 0.30\\
R33-37b	&16:33:02.1	&$-$24:37:17	&1999-Aug-09	& 0.26\\
~~~~--	&--		&--		&2000-Apr-11	& 0.22\\
R33-32	&16:33:17.0	&$-$24:32:16	&1999-Mar-09	& 0.34\\
~~~~--	&--		&--		&2000-Apr-10	& 0.18\\
~~~~--	&--		&--		&2000-Apr-12	& 0.30\\
R33-46	&16:33:22.3	&$-$24:46:16	&2000-Apr-12	& 0.30\\
R33-37a	&16:33:38.1	&$-$24:37:15	&2000-Apr-11	& 0.22\\
R34-28	&16:34:01.9	&$-$24:28:13	&1999-Mar-09	& 0.38\\
~~~~--	&--		&--		&1999-Aug-08	& 0.26\\
R34-37	&16:34:14.1	&$-$24:37:13	&2000-Apr-11	& 0.22\\
R34-38	&16:34:50.2	&$-$24:38:04	&1999-Aug-08	& 0.26\\
~~~~--	&--		&--		&1999-Aug-09	& 0.30\\
~~~~--	&--		&--		&2000-Apr-11	& 0.22\\ \hline
\end{tabular}
\end{center}
\end{table*}

The majority of studies of the $\rho$~Ophiuchi molecular cloud to date 
have concentrated on the L1688 sub-cloud, also known as the $\rho$~Oph 
main cloud. A number of filamentary clouds (L1709, L1740, L1744, L1755 
and L1765) extend from L1688 in a north easterly direction, and are often 
called the streamers or the cobwebs of Ophiuchus. To the southeast of L1688 
lies L1689, which also has filamentary clouds (L1712 and L1729) extending in 
roughly the same direction. \citet{1989ApJ...338..902L} measured masses 
for L1688 and L1689 of 1447 M$_\odot$ and 566 M$_\odot$ respectively, while 
the components of the streamers have lower masses of 100--300 M$_\odot$ each. 
The direction of the streamers turns to the south at the positions of both 
L1688 and L1689, forming a horseshoe shape in the case of L1688 and a 
boomerang shape in the case of L1689 (see Fig.~\ref{loren}).

Numerous studies of the L1688 cloud have revealed a rich history of star 
formation. Prompted by the association of early-type stars with the 
$\rho$~Ophiuchi dark cloud, \citet{1973ApJ...184L..53G} surveyed the cloud 
at 2~$\mu$m and discovered a large number of embedded young stars. Subsequent 
observations at a variety of wavelengths 
\citep[e.g.][]{1983ApJ...274..698W,1987AJ.....93.1182A,1989MNRAS.241..119W,
1989ApJ...340..823W,1994ApJ...434..614G,1995ApJ...439..752C,
1998A&A...336..150M,2000ApJ...545..327J,2004ApJ...611L..45J} 
have revealed large numbers of objects at every stage of star formation. 
Indeed, studies of this region have greatly increased our knowledge of the 
star-formation process as a whole.

\begin{figure*}
\includegraphics[angle=0,width=165mm]{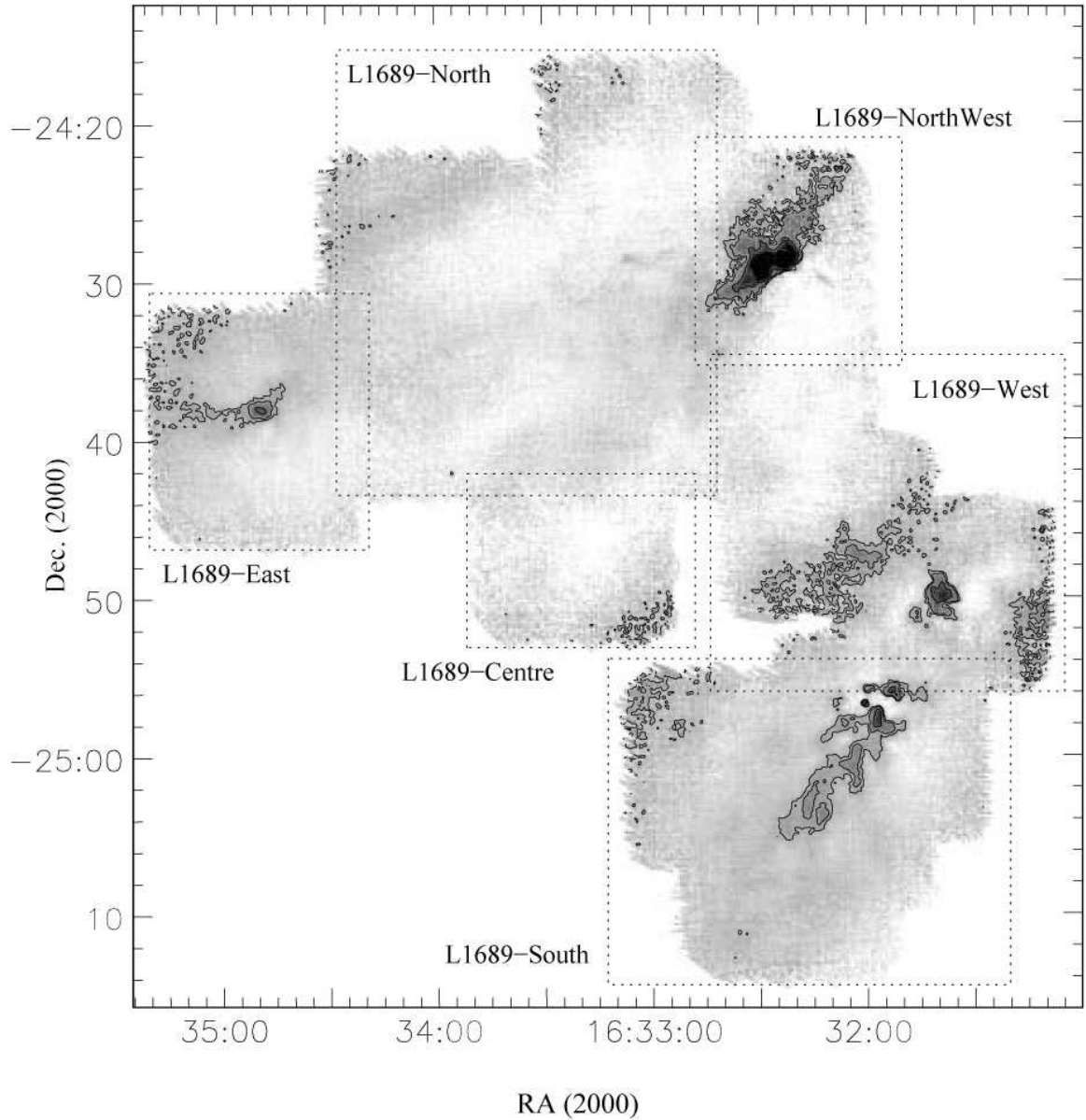}
\caption{The L1689 cloud at 850~$\mu$m. The contour levels are at 3$\sigma$, 
5$\sigma$, 10$\sigma$, 20$\sigma$, 40$\sigma$ and 80$\sigma$, where $\sigma$ 
is the mean noise level over the whole of the map, equal to 20 mJy/beam. 
Contours are based on the map smoothed to a resolution of 18 arcsec. The 
map regions in Table~\ref{noise_levels} are shown as dotted lines.}
\label{scan-map}
\end{figure*}

In contrast, the L1689 cloud appears to have relatively little star-forming 
activity, concentrated in a small number of isolated sources such as 
IRAS~16293-2422 \citep{1990ApJ...365..269L}. As such, we have labelled 
L1689 `the dog that didn't bark'. In this paper, we endeavour to find out 
why there is such an apparent discrepancy between the star-formation activity 
in L1688 and L1689 when they appear so similar in CO maps -- see 
Fig.~\ref{loren}. We report on submillimetre observations of L1689 and 
compare them with those taken at other wavelengths, in order to understand 
the similarities and differences between L1688 and L1689.

\begin{table*}
\begin{center}
\caption{Table showing the noise levels for each region of the L1689 maps. 
The position of each of the regions is illustrated in Fig.~\ref{scan-map}.}
\label{noise_levels}
\vspace{0.5cm}
\begin{tabular}{lcccc}\hline
	    &\multicolumn{2}{|c|}{Region Centre}  & 
\multicolumn{2}{|c|}{1$\sigma$ noise (Jy/beam)}   \\ \hline
Region Name 	& RA (2000)	& Dec. (2000)		& {850~$\mu$m}	&
{450~$\mu$m} \\ \hline
L1689-West	& 16:31:40 	& $-$24:50:00 		&  0.029
 		& 2.4		\\
L1689-South	& 16:32:10 	& $-$25:03:00 		&  0.016
 		& 0.51		\\
L1689-NorthWest	& 16:32:30 	& $-$24:35:00 		&  0.047
 		& 3.4		\\
L1689-Centre	& 16:33:20 	& $-$24:47:00 		&  0.034
 		& 1.3		\\
L1689-North  	& 16:33:30 	& $-$24:33:00 		&  0.023
 		& 0.87		\\
L1689-East	& 16:34:50 	& $-$24:38:00 		&  0.015
 		& 0.39		\\ \hline 
\end{tabular}
\end{center}
\end{table*}

\section{Observations}

The observations were carried out using the Submillimetre Common User 
Bolometer Array (SCUBA) on the James Clerk Maxwell Telescope (JCMT). 
This instrument takes observations at 450 and 850~$\mu$m simultaneously 
through the use of a dichroic beam-splitter. The telescope has a resolution 
of 8 arcsec at 450$~\mu$m and 14 arcsec at 850~$\mu$m.

Observations of L1689 were carried out over several nights between 1999 March 
and 2000 April using the scan-map observing mode as listed in 
Table~\ref{obs_table}. A scan-map is made by scanning the array across the 
sky. The scan direction is 15.5$^\circ$ from the axis of the array in order 
to achieve Nyquist sampling. The array is rastered across the sky to build 
up a map several arcminutes in extent.

Time-dependent variations in the sky emission were removed by chopping the 
secondary mirror at 7.8 Hz. Due to a scan-map being larger in size than the 
chop throw, each source in the map appears as a positive and a negative 
source. In order to remove this dual-beam function, each region is mapped 
six times, with chop throws of 30, 44 and 68 arcsec in both RA and Dec 
\citep{1995mfsr.conf..309E}. The dual-beam function is removed from each 
map in Fourier space by dividing each map by the Fourier transform of the 
dual-beam function, which is a sinusoid. The multiple chop-throws allow for 
cleaner removal of the dual beam function in Fourier space. The maps are them 
combined, weighting each map to minimise the noise introduced at the spatial 
frequencies that correspond to zeroes in the sinusoids. Finally the map is 
converted back into normal space, where it no longer contains the negative 
sources \citep{SURF}.

The submillimetre zenith opacity at 450 and 850$~\mu$m was determined using 
the `skydip' method and by comparison with polynomial fits to the 1.3~mm sky 
opacity data, measured at the Caltech Submillimeter Observatory 
\citep{2002MNRAS.336....1A}. The average zenith optical depth at 850 $\mu$m 
is listed in Table~\ref{obs_table} for each observation. The telescope 
pointing was checked at regular intervals throughout the nights using 
planets, secondary calibrators and standard pointing sources. 

The data were reduced in the normal way using the SCUBA User Reduction 
Facility \citep{SURF}. Calibration was performed using observations of 
the planet Uranus taken during each shift. We estimate that the absolute 
calibration uncertainty is $\pm$10\% at 850$~\mu$m and $\pm$30\% at 
450$\mu$m, based on the consistency and reproducibility of the calibration 
from map to map. 

\section{Results}

\begin{figure}
\includegraphics[angle=0,width=83mm]{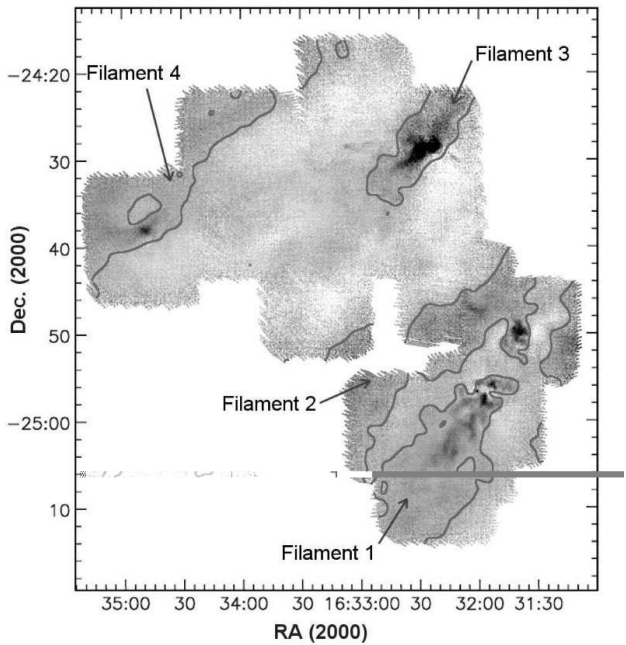}
\caption{Composite map showing the 850~$\mu$m scan-map of L1689 as a 
grey-scale. The superposed
contours are the 850-$\mu$m data smoothed to a 
resolution of 1 arcmin to highlight the large-scale structure. The names 
of the filaments are also marked (see text for details).}
\label{cobweb1}
\end{figure}

The map of L1689 is made up of a number of scan-maps, each of which is 12 
$\times$ 12 arcmin in size. The positions of the scan-maps were chosen to 
trace the $^{13}$CO emission mapped by \citet{1989ApJ...338..902L}. This is 
shown graphically in Fig.~\ref{overlay}, where the mapped area is overlaid 
in white on the black $^{13}$CO contours \citep{1989ApJ...338..902L}. The 
names and approximate positions of the different scan-map fields are shown 
in white. The background 0.5~$\mu$m image is taken from the Digitized Sky 
Survey \citep{1994IAUS..161..167L}, obtained using the {\em SkyView} interface
\citep{1994ASPC...61...34M}. The mapping strategy was to offset each scan-map 
from the neighbouring maps by approximately half a map-length. This was to 
avoid potential `striping' caused by the increased noise levels at the 
scan-map edges. The name of each scan-map is taken from the J2000 coordinates 
of the scan-map centre. The observation details are summarised in 
Table~\ref{obs_table}.

Fig.~\ref{scan-map} shows the data for the L1689 cloud at 850$~\mu$m.  
The area covered is approximately 0.5 deg$^2$, which at a distance of 128 pc 
is equal to $\sim$ 2.5 pc$^2$. As a result of the different weather conditions
at the telescope when the maps were made, and also the varying integration 
time per point, the noise level varies across the map. We have therefore 
split the map up into a number of regions, and measured the mean 1$\sigma$ 
noise level for each region. These are given in Table~\ref{noise_levels}. 
The extent of the different regions, and our names for them, are illustrated 
with dashed lines in Fig.~\ref{scan-map}.

A number of sources are seen in the 850~$\mu$m map. We have labelled these 
SMM~1$-$21 in order of increasing declination. The sources were identified 
by eye by selecting regions of flux density greater then 4-$\sigma$ above 
the background level. All the sources are smaller than $\sim 1-2$ arcmin. 
Any structure larger than this scale is deemed to be part of the filamentary 
structure of the cloud, and is discussed below. The identities of the 
relatively compact (i.e. $\leq 1$ arcmin) sources are discussed in the 
following section. The only source detected at 450~$\mu$m is the well-studied 
binary/multiple Class~0 protostar IRAS~16293-2422 
\citep{1989ApJ...337..858W,1992ApJ...385..306M,1993ApJ...402..655W}.

\begin{figure}
\includegraphics[angle=0,width=82mm]{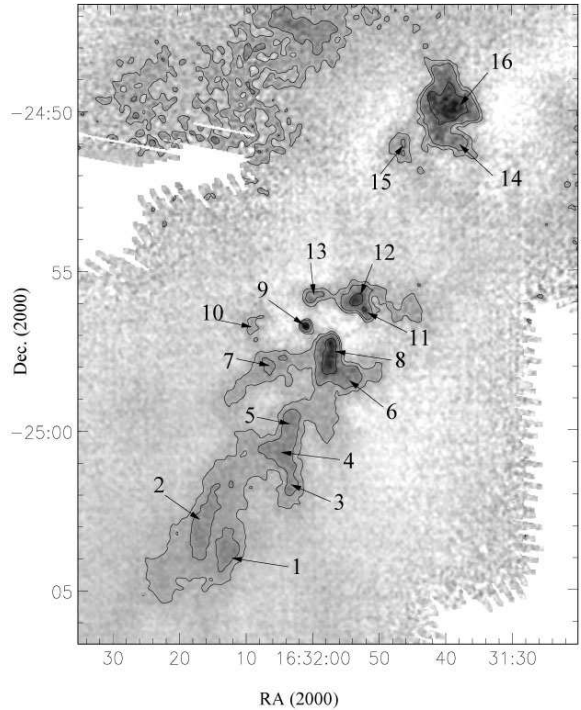}
\caption{An enlarged view of Filament~1 at 850~$\mu$m. The sources SMM$~1-16$ 
that are associated with this filament are labelled.}
\label{l1689smm}
\end{figure}

\begin{table*}
\begin{center}
\caption{The position and measured flux density of each source in the L1689 
scan-maps at 850~$\mu$m and 450~$\mu$m. Column~4 indicates whether or not the 
core is extended in the 14 arcsec 850~$\mu$m JCMT beam. The peak flux density 
is quoted for all sources (columns~7 and 9), with $3\sigma$ upper limits given
for undetected sources. The integrated flux density within an aperture is 
given for extended sources in columns~8 and 10. The semi-major \& semi-minor 
axes of these apertures are given in columns~5 and 6. The names given to 
these sources in other surveys are given in Table~\ref{masses}.}
\label{cores}
\begin{small}
\vspace{0.5cm}
\begin{tabular}{|l|c|c|c|c|c|c|r|c|c|}\hline
&&&& \multicolumn{2}{|c|}{Aperture} 	& \multicolumn{2}{|c|}{850~$\mu$m} & 
\multicolumn{2}{|c|}{450~$\mu$m} 			\\ \cline{7-10}
Source 	& RA 		& Dec.		& Ext.	& Semi-major& Semi-minor  &
Peak	&Int.	& Peak		& Int. 	\\
Name	&(2000)		&(2000)		& 	& (arcsec)  &(arcsec)
		&(Jy/beam)	&(Jy)	&(Jy/beam)	&(Jy)	\\ \hline
SMM~1	&16:32:12.5	&-25:03:53	&Y	&32	    &24		&0.15
	&0.75	&$<1.5$	&-- \\
SMM~2	&16:32:16.1	&-25:02:44	&Y	&67	    &29		&0.14
	&1.6	&$<1.5$	&-- \\
SMM~3	&16:32:02.7	&-25:01:48	&Y	&24	    &22		&0.15
	&0.50	&$<1.5$	&--\\
SMM~4	&16:32:04.1	&-25:00:53	&Y	&38	    &22		&0.18
	&1.1	&$<1.5$	&--\\
SMM~5	&16:32:03.0	&-24:59:48	&Y	&37	    &22		&0.15
	&0.67	&$<1.5$	&--\\
SMM~6	&16:31:54.6	&-24:58:22	&Y	&30	    &20		&0.22
	&0.94	&$<1.5$	&--\\
SMM~7	&16:32:06.3	&-24:58:01	&Y	&41	    &27		&0.11
	&0.49	&$<1.5$	&--\\
SMM~8	&16:31:57.5	&-24:57:40	&Y	&43	    &25		&0.30
	&1.4	&$<1.5$	&--\\
SMM~9	&16:32:01.0	&-24:56:44	&N	&13	    &13		&0.44
	&--	&$<1.5$	&--\\
SMM~10	&16:32:08.6	&-24:56:41	&Y	&39	    &27		&0.11
	&0.45	&$<1.5$	&--\\
SMM~11	&16:31:52.2	&-24:56:13	&Y	&17	    &15		&0.29
	&0.56	&$<1.5$	&--\\
SMM~12	&16:31:53.7	&-24:55:54	&Y	&36	    &21		&0.32
	&1.4	&$<1.5$	&--\\
SMM~13	&16:32:00.3	&-24:55:53	&Y	&31	    &21		&0.23
	&0.72	&$<1.5$	&--\\
SMM~14	&16:31:37.5	&-24:51:09	&Y	&36	    &26		&0.27
	&1.5	&$<7.1$	&--\\
SMM~15	&16:31:46.8	&-24:51:10	&Y	&34	    &27		&0.20
	&1.2	&$<7.1$	&--\\
SMM~16	&16:31:39.2	&-24:49:48	&Y	&63	    &51		&0.40
	&7.8	&$<7.1$	&--\\
SMM~17	&16:33:55.7	&-24:42:06	&N	&12	    &11		&0.21
	&--	&$<2.6$	&--\\
SMM~18	&16:34:48.4	&-24:38:04	&Y	&43	    &30		&0.24
	&2.0	&$<1.2$	&--\\
SMM~19	&16:32:29.3	&-24:29:13	&Y	&48	    &33		&1.64
	&12	&$<10$	&--\\
SMM~20	&16:32:23.0	&-24:28:40	&Y	&39	    &28		&15.9
	&28	& 80	& 180\\
SMM~21	&16:33:06.2	&-24:28:38	&Y	&27	    &19		&0.12
	&0.36	&$<2.6$	&--\\ \hline
\end{tabular}
\end{small}
\end{center}
\end{table*}

Table~\ref{cores} gives the measured flux density for each source detected in 
the map. The peak flux density is given in Jy/beam for each source at both 450
 and 850 $\mu$m. 3$\sigma$ upper limits are given for sources that are 
undetected at 450 $\mu$m. The integrated flux density is also given for 
sources that are extended in the 14 arcsec JCMT beam. The semi-major and 
semi-minor axes of the elliptical apertures used to measure these integrated 
flux densities are given in columns~5 and 6 of Table~\ref{cores}.

Taking the map as a whole, we see that the cloud appears to be composed of a 
number of approximately parallel filaments, that run the length of the map. 
These are highlighted in Fig.~\ref{cobweb1}, where the contours are
the data after it has been 
smoothed to a resolution of 1 arcmin to illustrate the large scale structure. 
A number of sources associated with these filaments are also detected. The 
nature of these sources is discussed below.

Filament~1 lies at the southern end of the map, and extends 
northwest-southeast (NW-SE) through the L1689-West and L1689-South regions. 
The filament appears to have fragmented into a number of objects, forming 
more than one object across its width in some cases. The degree of 
fragmentation appears to be greater at the north of the filament, where the 
majority of the sources detected in this filament are found. The sources at 
the southern end of the filament are also seen to be less centrally condensed.
Fig.~\ref{l1689smm} shows this filament in more detail. The 16 sources in the
filament have been labelled. Most of these objects are significantly extended
in the JCMT beam, and have a fairly complex structure. This filament contains
the majority of the sources that are seen in the L1689 molecular cloud.

\begin{figure*}
\begin{tabular}{c c}
\includegraphics[angle=0,width=78mm]{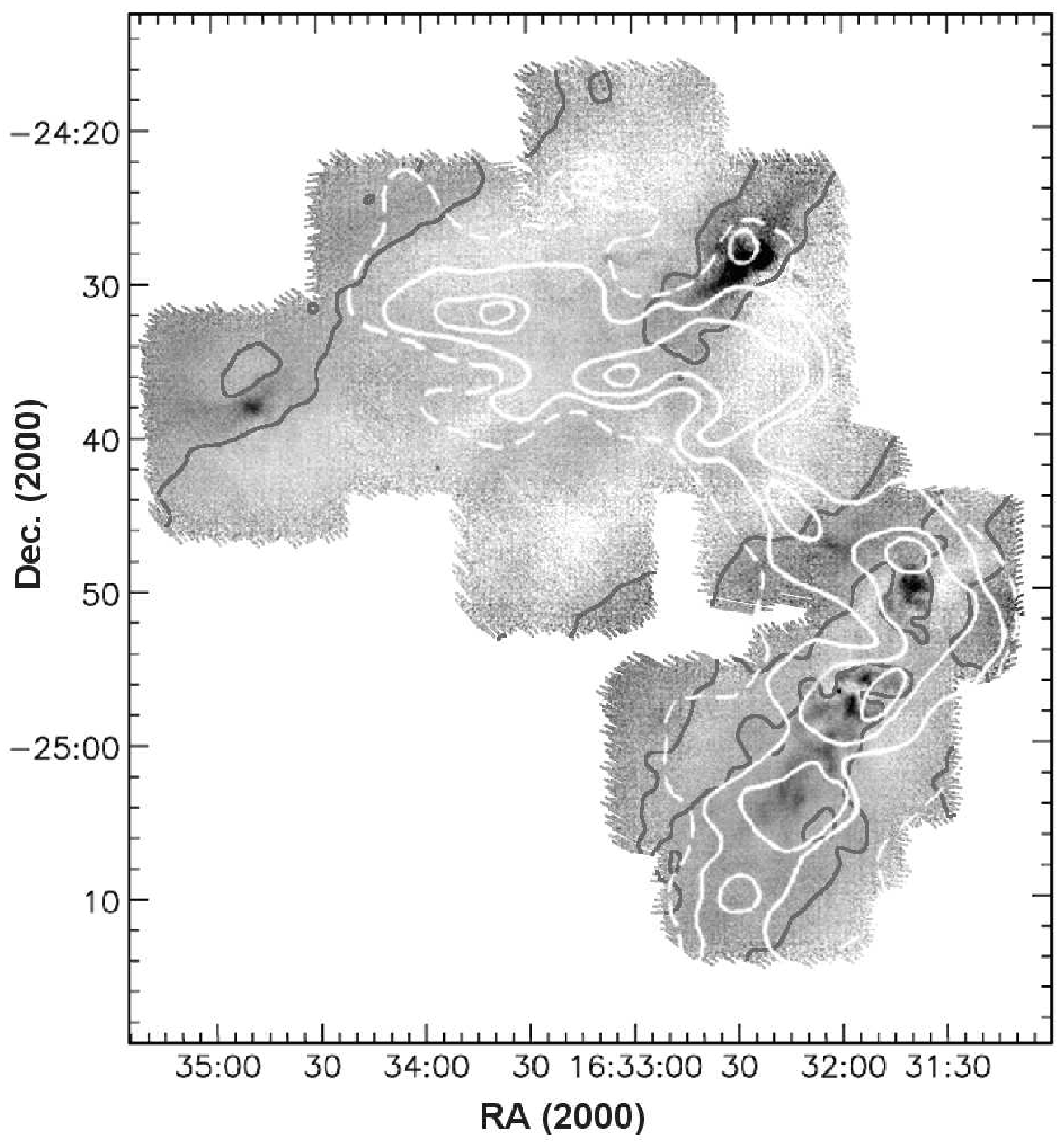} &
\includegraphics[angle=0,width=78mm]{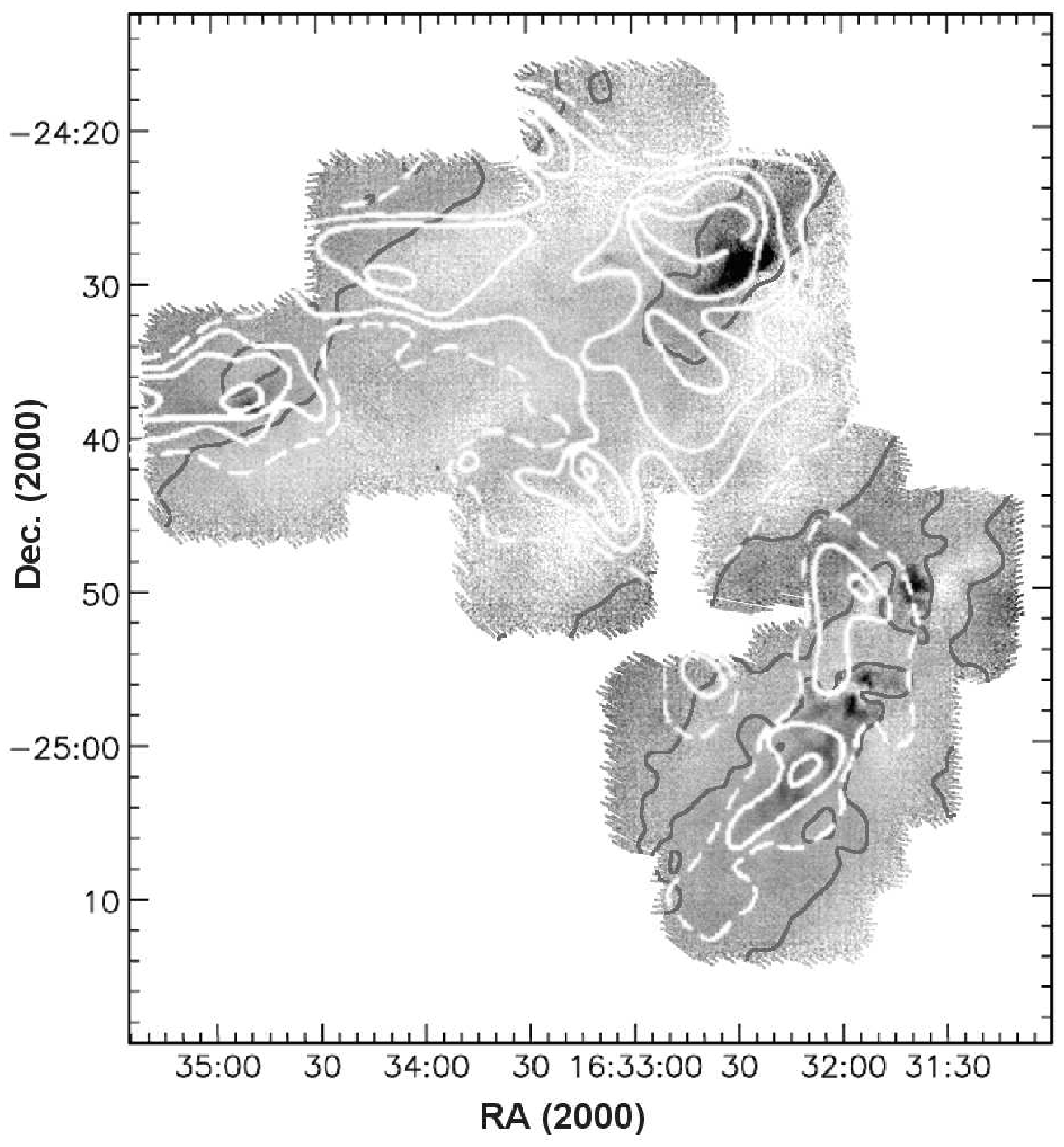} \\
(a) & (b) \\
\end{tabular}
\caption{Composite map showing the 850~$\mu$m scan-map of L1689 as a 
grey-scale. The dark contours are 850-$\mu$m data smoothed to a resolution 
of 1 arcmin to highlight the large-scale structure. The light contours are 
$^{13}$CO emission with ${\rm V_{SLR}}$ of (a) $4.68 - 5.02$ kms$^{-1}$ and 
(b) $3.32 - 3.66$ kms$^{-1}$ \citep{1989ApJ...338..902L}, and give antenna 
temperatures $T_R^\star(^{13}{\rm CO})$ of 2, 4, 6, 8, 10 and 12~K.
The continuum filaments can be seen at different velocities in the $^{13}$CO
data.}
\label{cobweb2}
\end{figure*}

Filament~2 runs roughly parallel to Filament~1 but slightly to the north, 
also in a NW-SE direction. There are no detected sources associated with 
this filament. The filament has been independently detected on a number of 
scan-maps. Unfortunately, Filament~2 has not been covered very well by the 
scan-mapping strategy. There is an area that has not been mapped in the 
middle of the filament, and most of the area that has been mapped has been 
covered only once. In addition, the filament has mostly been caught on the 
edges of the scan-maps, where the noise levels can be of order 1.5$\times$ 
higher. As a result of these factors, the level of noise along the filament 
is higher than average. We therefore suggest that this filament be re-mapped 
to confirm our findings.

Filament~3 extends from the L1689-NorthWest region, in a NW-SE direction also 
roughly parallel to Filament~1. The brightest source in this filament is the 
IRAS~16293 binary/multiple Class 0 protostar (labelled here SMM~20). There is 
also a weaker source 90 arcsec (approximately 14000 AU) to the southeast of 
IRAS~16293, named 16293E (SMM~19). It is interesting to note that the extended
emission surrounding IRAS~16293 lies parallel to the axes of the other 
filaments in the map. The filament is only detected in the northwest of the 
image, though if the line of the filament is extrapolated, a weak unresolved 
source (SMM~17) is detected at 
RA (2000) = 16$^{\rm h}$33$^{\rm m}$55.7$^{\rm s}$, 
Dec. (2000) = $-$24$^{\circ}$42$'$06$''$.

Filament~4 extends from the L1689-East region at the east of the map, in a 
NW-SE direction. Like Filament~2, it skirts along the edge of the map, 
therefore the noise is higher than average. The only object detected in the 
filament is the L1689B \citep{1983ApJ...266..309M} prestellar core 
\citep{1994MNRAS.268..276W,2005MNRAS.360.1506K} at 
RA (2000) = 16$^{\rm h}$34$^{\rm m}$48.6$^{\rm s}$, 
Dec. (2000) = $-$24$^{\circ}$38$'$00$''$, that we label here SMM~18. 
Filament~4 is the least completely mapped of the four filaments, but it 
is seen in more than one overlapping scan-map. We also note that the 
envelope of L1689B is elongated parallel with the filament axis. 

When observed with molecular line tracers, the $\rho$~Ophiuchi cloud is 
also filamentary in appearance \citep{1989ApJ...338..902L}. This filamentary 
structure is not restricted to the streamers that extend to the northeast, 
but is also seen in the larger L1688 and L1689 sub-clouds 
\citep{1983ApJ...274..698W,1989ApJ...338..902L}. 

To illustrate this, Figs~\ref{cobweb2}(a)\&(b) show our 850~$\mu$m data 
smoothed to highlight the filaments as in Fig.~\ref{cobweb1}. $^{13}$CO 
maps are overlaid as light contours \citep{1989ApJ...338..902L}. 
Figs.~\ref{cobweb2}(a)\&(b) show the $^{13}$CO radial velocity 
intervals $4.68 - 5.02$ kms$^{-1}$ and $3.32 - 3.66$ kms$^{-1}$ 
respectively. As can be seen, Filament~1 is clearly detected in 
$^{13}$CO at both of these velocity intervals. Filaments 3 and 4 
and the northern half of Filament~2 are seen at velocities 
3.32 -- 3.66 kms$^{-1}$ (Fig.~\ref{cobweb2}b).

\begin{table*}
\begin{center}
\caption{The prestellar core masses and protostellar envelope masses 
for the SCUBA 
detections in L1689. The assumed values of the temperature and the 
dust opacity for each of the sources is listed in columns 3 \& 4 (see text 
for details). The Bonnor-Ebert critical mass for the starless cores is given 
in column 6. The region of the cloud that each source is associated with is 
included in column~7. $^{\star}$IRAS~16293 is known to be a binary/multiple 
system. References in column~8 are as follows:
$^1$\citealp{2000A&A...361..555B}, 
$^2$\citealp{1994ApJ...434..614G}, 
$^3$\citealp{2001A&A...372..173B}, 
$^4$\citealp{2005MNRAS.360.1506K}, 
$^5$\citealp{1994ApJ...420..837A}, 
$^6$\citealp{1992ApJ...385..306M}, 
$^7$Castets et al. 2001, 
$^8$\citealp{1998MNRAS.300..733M}, 
$^9$\citealp{1996A&A...314..625A}.}
\label{masses}
\begin{small}
\vspace{0.5 cm}
\begin{tabular}{llcclccl}\hline
Source & Class 	& Temp	& $\kappa_{850}$& Mass 		& $M_{BE}$	& 
$^{13}$CO & Other 				\\ 
Name   &	& (K)   &(${\rm cm^2g^{-1}}$)&(M$_\odot$)&(M$_\odot$)	& 
region    & Names				\\ \hline
SMM~1  & ~Prestellar	& 12	& 0.01		& 0.3		&0.4
		& R55	    & L1689A$^4$			\\
SMM~2  & ~Prestellar	& 12	& 0.01		& 0.6		&0.7
		& R55	    &					\\
SMM~3  & ~Prestellar	& 12	& 0.01		& 0.2		&0.3
		& R55	    &					\\
SMM~4  & ~Prestellar	& 12	& 0.01		& 0.4		&0.4
		& R55	    &					\\
SMM~5  & ~Prestellar	& 12	& 0.01		& 0.3		&0.4
		& R55	    &					\\
SMM~6  & ~Prestellar	& 12	& 0.01		& 0.4		&0.4
		& R55	    &						\\
SMM~7  & ~Starless	& 12	& 0.01		& 0.2		&0.5
		& R55	    &					\\
SMM~8  & ~Prestellar	& 12	& 0.01		& 0.6		&0.5
		& R55	    & L1689SMM$^4$			\\
SMM~9  & ~I	& 30	& 0.02		& 0.02		&--		& 
R55	    & IRAS 16289-2450, L1689-IRS6$^2$, \\
	&	&	&		&		&		&
	    &ISO-209$^3$, L1689S-IRS67$^4$\\
SMM~10 & ~Starless	& 12	& 0.01		& 0.2		&0.5
		& R55	    &					\\
SMM~11 & ~II	& 30	& 0.03		& 0.02		&--		& 
R55	    & L1689-IRS5$^2$, ISO-204$^3$	\\
SMM~12 & ~Prestellar	& 12	& 0.01		& 0.6		&0.4
		& R55	    &					\\
SMM~13 & ~Prestellar 	& 12	& 0.01		& 0.3		&0.4
		& R55	    &					\\
SMM~14 & ~Prestellar	& 12	& 0.01		& 0.6		&0.5
		& R53	    &					\\
SMM~15 & ~Prestellar	& 12	& 0.01		& 0.5		&0.5
		& R53	    & R53$^1$				\\
SMM~16 & ~Prestellar	& 12	& 0.01		& 3.0 		&0.8
		& R53	    & R53$^1$				\\
SMM~17 & ~II	& 30	& 0.03		& 0.01		&--		& 
R65	    & RX J1633.9-2442$^8$		\\
SMM~18 & ~Prestellar	& 12	& 0.01		& 0.8		&0.5
		& R65	    & L1689B$^9$			\\ 
SMM~19 & ~0	& 30	& 0.02		& 0.6		&--
		& R57	    & IRAS 16293-2422E$^7$		\\
SMM~20 & ~0$^\star$& 30	& 0.02		& 1.3		&--		& 
R57	    & IRAS 16293-2422$^6$		\\
SMM~21 & ~Starless	& 12	& 0.01		& 0.1		&0.3
		& R57	    &					\\ \hline
\end{tabular}
\end{small}
\end{center}
\end{table*}

\section{Individual Sources}

\subsection{Identities of the detected sources}

In identifying the nature of submillimetre detected sources, there are a 
number of factors to take into account. To differentiate between a prestellar
 core and a Class 0 protostar, one can look for a centimetre radio source or a
 collimated molecular outflow, which would indicate the presence of a central 
hydrostatic object. Class 0 protostars are distinguished from more evolved 
classes of protostar by a large ratio of submillimetre to bolometric 
luminosity, or the presence of a centrally peaked and extended envelope 
\citep{2000prpl.conf...59A}. More evolved protostars (Class I and later) 
can be seen in the infrared. In this section, we discuss the identities of 
each of the sources, and summarise the source parameters in 
Table~\ref{masses}.

In order to estimate whether or not the sources are gravitationally bound, 
we compare their masses to the Bonnor-Ebert critical mass 
$M_{BE} = 2.4Ra_s^2/G$ \citep{1956MNRAS.116..351B,1955ZA.....37..217E}, 
where $R$ is the radius of the source (taken to be the geometric mean of 
the semi-major and semi-minor axes given in Table~\ref{cores}), $a_s$ is 
the isothermal sound speed, and $G$ is the gravitational constant.

The source SMM~1 corresponds to the prestellar core L1689A, which has 
previously been mapped in the submillimetre \citep{2005MNRAS.360.1506K} 
and also in the far-infrared \citep{2002MNRAS.329..257W}. This core's 
position, close to the leading edge of the L1689 cloud nearest the Upper 
Scorpius OB association (see Section~\ref{sequential}), could potentially 
explain the unusual temperature profile measured in the far-infrared by 
\citet{2002MNRAS.329..257W}.

The sources SMM~2-5 are all fairly extended cores that are not very 
centrally peaked. Their masses are comparable with the Bonnor-Ebert 
critical mass, as given in Table~\ref{masses}. These sources have not 
been observed previously, with the exception of SMM~5 which was detected 
in the millimetre \citep{1994ApJ...420..837A}, and possibly corresponds to 
the source VLA~16289, though the nature of this object is unknown. SMM~7 is 
also extended, though its mass is less than half of its Bonnor-Ebert
critical mass. Therefore this source may well not be bound, but a transient 
object. Hence we label it as starless rather than prestellar
in Table~4. This follows the distinction of definitions
made by \citet{PPV}, who stated that the difference
between a starless core and a prestellar core is that a prestellar
core should be consistent with being gravitationally bound.

SMM~8 is the prestellar core L1689SMM, which was mapped by 
\citet{2005MNRAS.360.1506K} using SCUBA in jiggle-map mode. 
There is no evidence of a protostar at the centre. The core is 
elongated in a north-south direction, though shows no evidence of 
fragmenting along its length. Both the morphology and flux density 
of our data are consistent with the earlier jiggle-map. There is a 
weaker source (SMM~6) to the south-west of SMM~8, which partially 
overlaps it.

SMM~9 is the Class I protostar IRAS 16289-2450. It has also been 
detected at a number of wavelengths from the near-infrared 
(L1689-IRS6 -- \citealp{1994ApJ...434..614G}), the mid-infrared 
(ISO-209 -- \citealp{2001A&A...372..173B}) and the millimetre 
(L1689S-IRS67 -- \citealp{1994ApJ...420..837A}). This object is 
centrally condensed in the submillimetre and no extended emission 
is detected. 

There is an abrupt gap in the filament between SMM~8 and the group of 
objects to the north (SMM~11, 12 and 13). This could be caused by the 
gravitational collapse of the filament onto the surrounding cores, or 
the cavity could have been cleared by outflows from IRAS 16289-2450 
(SMM~9), which lies within the cavity.

SMM10 is similar in appearance to its neighbour, SMM~7, and also appears 
not to be bound. SMM~11 is a Class II protostar which has previously been 
identified in the near-infrared (L1689-IRS5 -- \citealp{1994ApJ...434..614G}) 
and mid-infrared (ISO-204 -- \citealp{2001A&A...372..173B}). Approximately 25 
arcsec north of this (0.015 pc) lies SMM~12, which has not been previously 
observed at other wavelengths. This object is assumed to be prestellar in 
nature. SMM~13 has also not been detected at other wavelengths. Its mass is 
$\sim$70\% of the critical Bonnor-Ebert mass, therefore it may not be bound.
Again we label this source as starless in Table~4.

SMM~16 lies at the northern end of Filament~1, and is one of the brighter 
sources in the map. It appears to have some internal structure, though this 
is poorly resolved in the 14 arcsec beam. This suggests that SMM~16 could be 
composed of multiple unresolved sources, surrounded by a common envelope.  
The diameter of the source is approximately 2 arcmin (0.08 pc at a distance 
of 130 pc). A VLA survey has been conducted of this region 
\citep{1988AJ.....96.1394S}. However, no centimetre sources coincident with 
SMM~16 were detected. This indicates that the source (or sources) may be 
prestellar in nature.

Two sources (SMM~14 \& 15) are located 1--2 arcmin from the centre of SMM~16. 
Again there is no evidence of associated protostars 
\citep{1988AJ.....96.1394S}, we therefore assume that these objects are 
prestellar, based on their Bonnor-Ebert masses. 
The three sources SMM~14, 15 and 16 are coincident with the 
starless core R53 which has been observed in absorption in the mid-infrared 
\citep{2000A&A...361..555B}.

SMM~17 is a very weak isolated unresolved source, which lies 18 arcsec from 
the Class~II YSO RX~J1633.9-2442, which was identified through optical 
spectroscopy follow-up observations to a ROSAT X-ray survey of $\rho$~Ophiuchi
\citep{1998MNRAS.300..733M}. It is assumed that the submillimetre emission is 
associated with this source.

SMM~18 is the well studied prestellar core L1689B \citep[e.g.][and references
therein]{2005MNRAS.360.1506K}. This was one of the original starless ammonia 
cores mapped by \citet{1983ApJ...266..309M} and subsequently determined to be 
prestellar in nature \citep{1994MNRAS.268..276W}. Subsequent mapping with the
1.3 mm MPIfR bolometer array at the IRAM 30m telescope confirmed the results of 
the earlier survey \citep{1996A&A...314..625A}, and provided a more detailed 
picture of the density profile of the core. The morphology of L1689B, as shown
in Fig.~\ref{scan-map}, is elongated in a roughly east-west direction. This 
is consistent with the 1.3 mm map \citep{1996A&A...314..625A} and also a 
SCUBA jiggle-map of the core \citep{2005MNRAS.360.1506K}. 
\citet{2002MNRAS.329..257W} performed a greybody fit to the SED of L1689B 
between 90~$\mu$m and 1.3 mm using ISO, JCMT and IRAM data, and found a 
best-fit temperature of $11.9^{+0.7}_{-0.5}$~K. 

The brightest source in the map is SMM~20, which is the binary/multiple 
Class 0 protostar IRAS~16293. This object has been well studied at a number 
of wavelengths \citep[e.g.][]{1989ApJ...337..858W,1992ApJ...385..306M,
1993ApJ...402..655W,1998A&A...331..372C,2001A&A...375...40C,
2004A&A...418..607C}, and is used as a secondary calibrator for 
submillimetre observations using SCUBA \citep{1994MNRAS.271...75S}. 
It appears to be a hierarchical system, with two components (16293A1 \& A2) 
separated by 47 AU, and a third source (16293B) located approximately 800 AU 
to the northwest \citep{1989ApJ...337..858W}. The objects are Class 0 
protostars and are the sources of a quadrupolar molecular outflow, though 
16293B shows no evidence of current outflow activity 
\citep{1993ApJ...402..655W}. 16293A and B are separated by 5 arcsec and are 
therefore unresolved in the 14 arcsec JCMT beam. Molecular line studies of 
species that adsorb to grain surfaces 
\citep[e.g.][]{2001A&A...375...40C,2004A&A...413..609W} indicate that the 
envelope of IRAS~16293 is composed of a hot core, approximately 2 arcsec 
(260 AU) across, at a temperature of $\sim 100$~K. The core is surrounded 
by a cooler outer envelope at a temperature of 20 -- 30~K. 

SMM~19 corresponds to 16293E, which is located approximately 85 arcsec 
(0.05 pc) to the east of IRAS~16293. \citet{2001A&A...375...40C} have 
argued that 16293E is a Class 0 protostar, on the grounds of a detected 
molecular outflow. \citet{2004ApJ...608..341S} did not detect 
the outflow from 16293E, and argued that the source is prestellar.
This may have been due to a sensitivity effect, so
we treat this source as a Class~0.

SMM~21 is a weak source with no previous detections at other wavelengths. 
Comparison with its Bonnor-Ebert critical mass indicates that it is probably 
not bound. We label it as a starless core.

\subsection{Source Masses}\label{source_masses}

Table~\ref{masses} summarises the results of the previous section and gives 
mass estimates for all of the detected sources. The envelope mass $M$ is 
calculated using:

\begin{equation}
M=\frac{S_{850}D^2}{\kappa_{850}B_{850,T}},
\end{equation}

\noindent
where $S_{850}$ is the 850~$\mu$m flux density, $D$ is the distance to the 
source, $\kappa_{850}$ is the mass opacity of the gas and dust, and 
$B_{850,T}$ is the Planck function at temperature $T$. 

For the starless and prestellar cores in this sample, we assume a value for 
$\kappa_{1.3mm}$ of ${\rm 0.005~cm^2g^{-1}}$, as recommended by 
\citet{1995P&SS...43.1333H}, and scale this to 850~$\mu$m using a canonical 
value of $\beta$ of 1.5 \citep{1994ApJ...420..837A}, giving $\kappa_{850\mu m}
= 0.01~\rm{cm^2g^{-1}}$. These values of $\kappa$ assume a gas-to-dust ratio 
of 100 \citep{1983QJRAS..24..267H}. For the denser envelopes around Class 0 
and I protostars, it is assumed that the formation of icy mantles will 
increase the $\kappa_{1.3mm}$ by a factor $\sim 2$ \citep{1994ApJ...420..837A,
1995P&SS...43.1333H}. Again, a canonical $\beta$ of 1.5 is used, giving a 
value of $\kappa_{850\mu m} = 0.02~\rm{cm^2g^{-1}}$ for these classes of 
object. For the higher density circumstellar disks surrounding Class II YSOs, 
grain agglomeration will increase $\kappa_{1.3mm}$ still further 
\citep{1995P&SS...43.1333H}. \citet{1990AJ.....99..924B} found that the best 
fits to the submillimetre spectra of T-Tauri stars are given by 
$\kappa_{1.3mm} = 0.02~\rm{cm^2g^{-1}}$ and $\beta = 1$. These values 
correspond to $\kappa_{850\mu m} = 0.03~\rm{cm^2g^{-1}}$.

The temperature of 12~K calculated for L1689B \citep{2002MNRAS.329..257W} 
was used for all of the prestellar and starless cores. For protostellar envelopes, the volume averaged 
dust temperature is assumed to be 30~K \citep{1994ApJ...420..837A}. 

\section{Comparison of L1689 with L1688}\label{5:l1689:comparison}

\subsection{Cloud constituents}

It is interesting to compare the L1688 and L1689 clouds, as they share the 
same environment, yet previous studies have indicated that L1688 has a much 
higher level of star formation \citep[e.g.][]{1990ApJ...365..269L}. 
\citet{1989ApJ...338..902L} measured the mass of L1688 and L1689 from 
$^{13}$CO maps to be 1447 and 566 M$_\odot$ respectively, differing by a 
factor of 2.6. The area on the sky of L1688 (measured at the 6~K contour of 
the $^{13}$CO maps --- see Fig.~\ref{loren}) is 1.4 times larger than L1689. 
Assuming that the two clouds have approximately the same aspect ratio (which 
is a fair assumption, given their similar morphology), the volume of L1688 is 
1.7 times that of L1689. The average density of L1688 is therefore 
approximately 1.5 times larger than that of L1689. As a result of this, 
we might expect star formation in L1688 to proceed more rapidly than in 
L1689.

\begin{figure}
\includegraphics[angle=0,width=83mm]{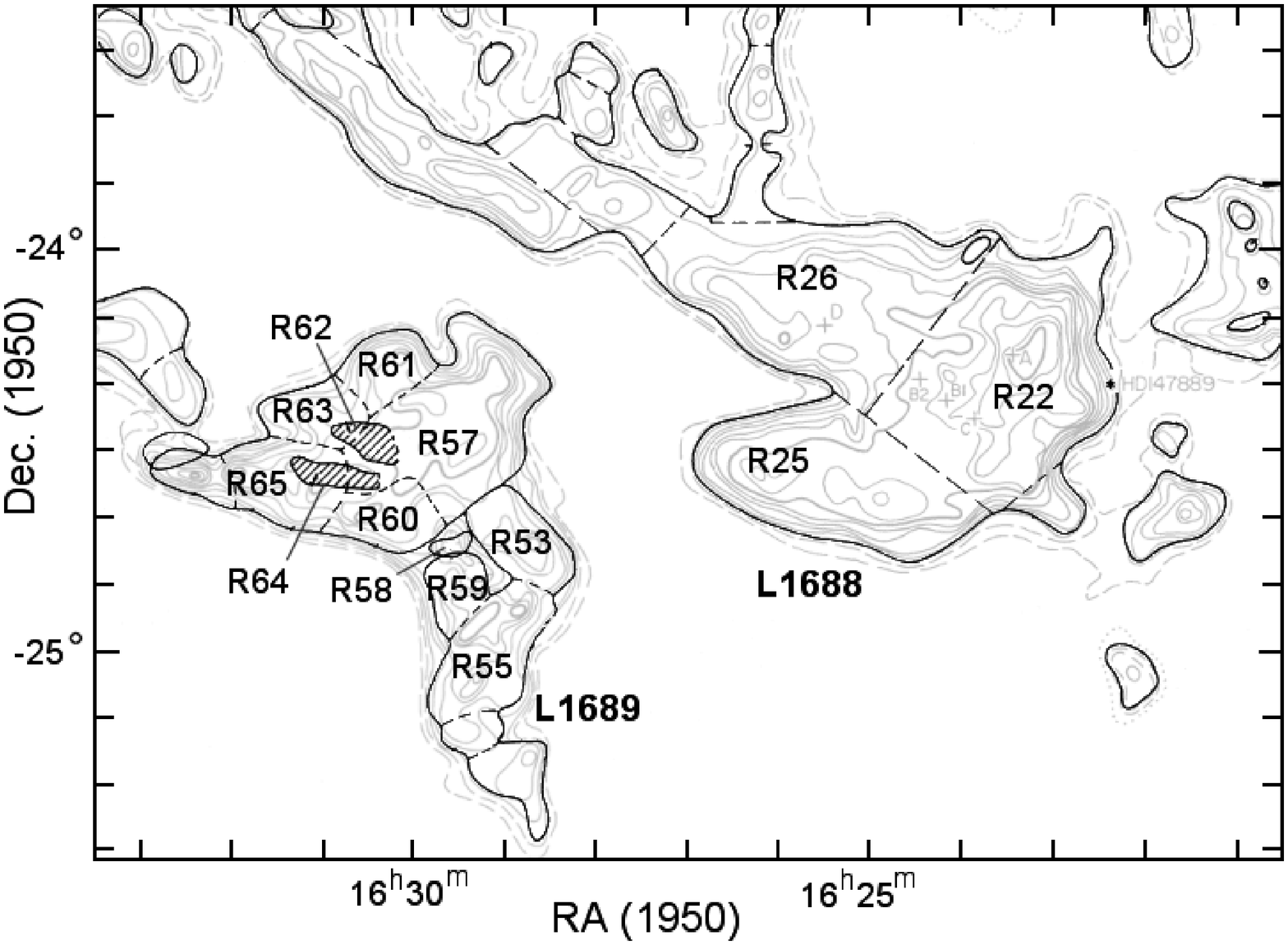}
\caption{The different regions of L1688 and L1689, defined by spatial or 
velocity separations \citep{1989ApJ...338..902L}. The region names are 
marked for areas of the clouds which have been mapped in the submillimetre.}
\label{l1689_regions}
\end{figure}

\citet{1989ApJ...338..902L} divided the two clouds into 89 different regions, 
which have a clear spatial or velocity separation. These regions are used in 
the following analysis to compare the star-formation activity across the 
clouds. Fig.~\ref{l1689_regions} shows the region boundaries overlaid on 
the $^{13}$CO contour map. The names of the regions that have been mapped 
in the submillimetre are marked on Fig.~\ref{l1689_regions} (see below).

\citet{1990ApJ...365..269L} investigated differences in the star-formation 
rates of the different regions of the $\rho$~Ophiuchi cloud. To do this, 
they compared the total luminosity ($L$) of the young stars associated with 
each region, with the mass ($M$) of the gas, determined using the $^{13}$CO 
measurements \citep{1989ApJ...338..902L}. They measured values of $L/M$ of 
$5.2~L_{\odot}/M_{\odot}$ for L1688 and $0.1~L_{\odot}/M_{\odot}$ for L1689, 
indicating a much larger star-formation efficiency (SFE) in L1688. This 
method of estimating the SFE is biased towards high mass stars. 
\citet{1990ApJ...365..269L} also calculated the ratio $N/M$, where $N$ is 
the number of young stars associated with each region. This estimate of the 
SFE is biased towards low mass stars. They again found that of all of the 
components of $\rho$~Ophiuchi, L1688 has the highest SFE, with a value of 
$N/M$ of 0.06. However, the value of $N/M$ of 0.05 for L1689 is only slightly 
lower.

These methods are both probes of the past star-formation activity in the two 
clouds. In the following section, we look at the mass of the material 
currently involved in star formation within each region of the two clouds, 
in order to compare the current star-formation activity. 

\begin{table}
\begin{center}
\caption{The masses measured using different tracers for the L1688 and L1689 clouds. The mass of each $^{13}$CO region is given in column~3 \citep{1989ApJ...338..902L}. The sum of the masses of all starless and prestellar cores in each of the regions is given in column~4 and simply labelled $M_{\rm pre}$. These were calculated using the SCUBA data from this study for L1689, and SCUBA data from \citet{2000ApJ...545..327J}, as well as IRAM 30m data from  \citet{1998A&A...336..150M}, for L1688. All the masses quoted in the table have been scaled such that they correspond to an assumed distance of 130~pc and temperature of 20~K. For each ${\rm ^{13}CO}$ region, the ratio of the prestellar and starless mass to the total mass of the region is given in 
column~5. The prestellar mass for R26 is calculated using table 2 of \citet{1998A&A...336..150M}. Note the difference of a factor of two between the total M$_{\rm pre}$/M$_{\rm ^{13}CO}$ for L1688 and L1689.}
\label{table_comparisons}
\begin{small}
\vspace{0.5 cm}
\begin{tabular}{|c|c|c|c|c|}\hline
Cloud & ${\rm ^{13}CO}$ & $M_{\rm ^{13}CO}$ & 
$M_{\rm pre}$ & $\underline{M_{\rm pre}}$ \\
Name   	& Region	& (M$_\odot$)	    & 
(M$_\odot$)   & $M_{\rm ^{13}CO}$  \\ \hline
L1688 	& R22 		& 557 	& 16.4	& 0.029	\\ 
	& R25		& 243	& 0.6	& 0.002	\\
	& R26		& 136	& 0.2	& 0.001	\\
{\bf Total}	& 	& {\bf 936}	& {\bf 17.2}	& {\bf 0.018}	\\ \hline
L1689	& R53		& 40	& 1.3	& 0.033	\\ 
 	& R55 		& 55 	& 1.4 	& 0.025	\\ 
 	& R57 		& 133 	& 0.1 	& 0.0008 \\ 
	& R58		& 1	& 0	& 0	\\
	& R59		& 15	& 0	& 0  	\\
	& R60		& 20	& 0	& 0  	\\
	& R61		& 7	& 0	& 0  	\\
	& R62		& 7	& 0	& 0  	\\
	& R63		& 13	& 0	& 0  	\\
	& R64		& 4	& 0	& 0  	\\
 	& R65 		& 26 	& 0.3 	& 0.012 	\\ 
{\bf Total}	& 	& {\bf 321}	& {\bf }	& {\bf 0.010}	\\ 
\hline
\end{tabular}
\end{small}
\end{center}
\end{table}

\subsection{Star-formation efficiency}

\citet{2000ApJ...545..327J} mapped the densest parts of the L1688 cloud 
using the SCUBA camera on the JCMT. Together with the data presented here, 
these make an ideal dataset to compare the current star-formation activity of 
the two clouds. \citeauthor{2000ApJ...545..327J} smoothed their map of L1688 
by twice the largest chop throw (130 arcsec), and subtracted this from the 
map. This was done in order to remove all large scale structure, some of 
which could result from residual atmospheric noise or incorrectly fitted 
baselines in the data reduction process. 

For the purposes of directly comparing the two datasets, we carried out the same step, and subtracted 
the large scale structure. In the following analysis, all the core masses 
are calculated assuming a distance to the region of 130~pc, 
and a dust temperature of 20~K. Again, this is to allow a direct comparison 
of the two data-sets.

The sensitivities of the two surveys are not significantly different. The 
SCUBA map of L1688 has a $1 \sigma$ sensitivity of approximately 10 mJy/beam 
\citep{2000ApJ...545..327J}, compared to our map of L1689, which has a 
$1 \sigma$ sensitivity of 15 -- 30 mJy/beam (see Table~\ref{noise_levels}). 

The SCUBA map of L1689 presented here covers most of the $^{13}$CO regions 
named by \citet{1989ApJ...338..902L} between R53 and R65 (see Table 
\ref{table_comparisons} and Figure \ref{l1689_regions}). 
The area covered by the SCUBA map of L1688 is approximately coincident with R22. 
The SCUBA map also covers the portion of R25 that is coincident with 
$\rho$-Oph~F \citep[see][]{1986ApJ...306..142L,1990ApJ...365..269L}. 
Though this region was not completely mapped by \citet{2000ApJ...545..327J}, 
a wider area survey of $\rho$ Ophiuchus \citep{2004ApJ...611L..45J} has 
indicated that there is very little star-formation activity in R25 except 
that contained in $\rho$-Oph~F.

The SCUBA map of L1688 does not include the $\rho$-Oph~D core in R26, 
therefore data obtained using the IRAM 30m telescope \citep{1998A&A...336..150M} 
were used to extend the following analysis to this region. 
\citeauthor{1998A&A...336..150M} separated the IRAM data into large and small 
scale structure using a wavelet analysis. We use the core masses for their 
small scale `clumps' (see their table 2), which, like the 
\citet{2000ApJ...545..327J} data, have effectively had the larger scale 
structure removed. For consistency with the other data-sets, we also 
scale the R26 core masses to a temperature of 20~K and a distance of 130~pc.

Table~\ref{table_comparisons} gives the mass of each cloud region, 
measured from the $^{13}$CO emission \citep{1989ApJ...338..902L}. 
The total mass of prestellar and starless cores within each region ($M_{\rm pre}$) 
is given in column~4. This is calculated from the submillimetre dust 
emission from the cores in L1688 
\citep{1998A&A...336..150M,2000ApJ...545..327J} and L1689. The cores 
that contain protostellar sources have not been included in these values. 
This is so we can evaluate and compare the potential yield of the next 
generation of star formation in the two clouds.

Table~\ref{table_comparisons} shows that the current star-formation activity 
is highly variable across both clouds. R22 has the largest mass of dense 
cores, with 16 M$_\odot$ compared to $< 2$ M$_\odot$ in each of the 
regions of L1689. This may be reasonably expected, as the R22 region is 
significantly more massive than the other regions. In order to compare the 
fraction of the
mass in dense cores in the different regions directly, we divide the mass in 
cores by the mass of the region, measured from $^{13}$CO maps
\citep{1989ApJ...338..902L}. This is given for each region in column~5 of Table~\ref{table_comparisons}.

When the two clouds (L1688 and L1689) are considered as a whole, L1688 has a
much higher fraction (by a factor of 2) of its mass contained in dense cores, with a value of $M_{\rm pre}/M_{\rm ^{13}CO}$ of 0.018, compared to 0.010 for L1689.
Even if we ignore the regions of L1689 that contain no submillimetre cores, and
only consider R53, R55, R57 \& R65, the total prestellar core mass ratio
for L1689 only rises to 0.012. This is still only two-thirds of 
the value seen in L1688. This indicates that L1688 will probably 
have a higher star formation efficiency 
in the next generation of star formation. 

However, the values of $M_{\rm pre}/M_{\rm ^{13}CO}$ for the most active regions of the two clouds (i.e. R22 in L1688 and R53 \& R55, which make up Filament 1 in L1689) are in fact similar (although the absolute masses are much smaller in R53 and R55). This could be indicating that parts of L1689 may not be as infertile as was previously thought. We note that these three regions are located at the western edges of L1688 and L1689. We consider the significance of this in the following section.

We finally note that the fraction of prestellar material in  $\rho$-Oph~A \citep[see][]{1986ApJ...306..142L,
1990ApJ...365..269L} is an order of magnitude higher than in the rest of 
R22 \citep[in agreement with][]{1998A&A...336..150M}. To calculate this, we compared the mass of compact submillimetre condensations in 
$\rho$-Oph~A \citep{2000ApJ...545..327J} with the mass measured using 
C$^{18}$O \citep[see][their table 2 and fig.~4]{1983ApJ...274..698W}. 

To summarise this section, we have found that on average, the L1688 cloud 
has a much higher fraction of its mass in dense cores than L1689. The 
western edges of both L1688 (R22) and L1689 (filament 1) have comparable
fractions. 
$\rho$-Oph~A is significantly more efficient at converting its mass into 
dense cores than any other component of the molecular cloud, and it appears 
likely that the existing young star cluster will continue to grow in richness. 
The regions to the east of both clouds have a much lower mass fraction in 
dense cores.

\subsection{Sequential star formation}\label{sequential}

One possible explanation for the picture painted in the previous section is 
that the star formation in both of the clouds is being affected by an external
influence, such as the by-products of nearby young massive stars. This 
scenario is known as sequential or triggered star formation, and was first 
suggested by \citet{1977ApJ...214..725E}. 

In this picture, the stars which 
have already formed, interact with the molecular cloud, triggering further 
star formation, which propagates through the cloud. This scenario has 
previously been used to explain the star formation in $\rho$~Oph 
\citep[e.g.][]{1986ApJ...306..142L,1989ApJ...338..902L}. In this case, 
we believe that the star formation in both L1688 and L1689 is being 
influenced by members of the Upper Scorpius OB association. 

The most massive and luminous nearby component of the Upper Scorpius OB 
association is $\sigma$~Sco. This is a hierarchical multiple containing 
an O9V star and a B2III star and at least two other B stars 
\citep[e.g.][]{1992A&A...261..203P}. $\sigma$~Sco lies at a distance of 
100 arcmin from L1688 (projected distance 4~pc), and 150 arcmin from L1689 
(projected distance 6~pc). It appears to be a prime candidate for the 
triggering mechanism in both of these clouds.

This hypothesis is consistent with the observation that the majority of the star formation in both clouds is occurring primarily in regions R22, R53 \& R55, which make up parallel filaments at the edge of each cloud closest to $\sigma$~Sco. These filaments are perpendicular to the line of sight joining each cloud and $\sigma$~Sco. In addition, the extensions of both L1688 and L1689 to the north-east lie on a line directly away from $\sigma$~Sco \citep{1989ApJ...338..902L}. In this hypothesis, the eastern regions of the two clouds are shielded from the triggering source by the two filaments. We note that R57 is unusual in that it is located on the western edge of L1689, yet has a very small value of $M_{\rm pre}/M_{\rm ^{13}CO}$. However, R57 has recently formed the protostellar sources SMM 19 \& 20, and if the mass associated with these sources is included, then the fertility of R57 becomes comparable with the regions of Filament 1. 

The influence of $\sigma$~Sco has clearly had a different effect on the two 
clouds, as is evidenced by the young proto-cluster that has formed in L1688, 
and the lack of such a cluster in L1689. We hypothesise that this difference 
in star-formation activity in the two clouds is simply due to the fact that 
L1688 is closer to $\sigma$~Sco. Hence it suffers a greater effect according 
to the square of the distance difference.

The sequential star formation in this region can be traced back a number of 
generations. The Upper Scorpius OB association is adjacent to an older OB 
association called Upper Centaurus-Lupus. \citet{2001ASPC..243..791P} suggest 
that the massive stars in Upper Centaurus-Lupus triggered the star formation 
in Upper Scorpius, just as Upper Scorpius is triggering star formation in 
$\rho$~Oph. It appears though that this is the end of the line for sequential 
star formation, as the only massive star to have formed in L1688 or L1689 is 
the young B3 star S1 \citep{1973ApJ...184L..53G}. It is therefore 
unlikely that this star will have the energy required to propagate the star 
formation as effectively
through the remainder of the $\rho$~Ophiuchi cloud in the same way.

\section{Conclusions}

We have mapped the L1689 cloud at submillimetre wavelengths, and detected a 
number of cores and protostellar envelopes, some of which are reported here 
for the first time. We have detected filaments in the submillimetre continuum 
that are also seen in the $^{13}$CO observations of the cloud. All of the 
detected star-formation activity in L1689 is contained within these filaments,
the majority being in Filament~1. 

We have compared the potential
future rate of star formation of L1689 with that of L1688 by looking at the 
fraction of
mass of each cloud that is currently in starless and prestellar cores. We find that when 
normalised to the total mass of each cloud, the L1688 cloud has on average
a much higher percentage of mass in cores. 
This is consistent with the higher 
star-formation activity observed in L1688.

We deduce that L1688 and L1689 are both examples of 
triggered star formation caused by $\sigma$~Sco and the reason that the 
L1689 dog has not barked is that it is less massive
than L1688 and is
further away from $\sigma$~Sco.

\section*{Acknowledgements}

The authors wish to thank the referee, Doug Johnstone, for comments which improved this manuscript. The authors would also like to thank the staff of the JCMT for assistance with the observations. The JCMT is operated by the Joint Astronomy Centre, Hawaii, on behalf of the UK PPARC, the Netherlands NWO, and the Canadian NRC. DJN acknowledges PPARC for PDRA support.

The Digitized Sky Survey was produced at the Space Telescope Science Institute
under U.S. Government grant NAG W-2166. The images of these surveys are based 
on photographic data obtained using the Oschin Schmidt Telescope on Palomar 
Mountain and the UK Schmidt Telescope. The plates were processed into the 
present compressed digital form with the permission of these institutions.

The authors acknowledge the use of NASA's {\em SkyView} facility 
(http://skyview.gsfc.nasa.gov) located at NASA Goddard Space Flight Center. 
The authors also gratefully
acknowledge R. Loren for permission to use the $^{13}$CO maps in 
Figures 1, 2, 6 and 7.


\begin{thebibliography}{}

\bibitem[\protect\citeauthoryear{Andr\'{e} \& Montmerle}{1994}]
{1994ApJ...420..837A} 
Andr\'{e} P., Montmerle T., 1994, ApJ, 420, 837 

\bibitem[\protect\citeauthoryear{Andre, Montmerle, \& Feigelson}{1987}]
{1987AJ.....93.1182A} 
Andre P., Montmerle T., Feigelson E.~D., 1987, AJ, 93, 1182 

\bibitem[\protect\citeauthoryear{Andr{\' e}, Ward-Thompson, \& Barsony}
{Andr{\' e} et al.}{1993}]{1993ApJ...406..122A} 
Andr{\' e} P., Ward-Thompson D., Barsony M., 1993, ApJ, 406, 122 

\bibitem[\protect\citeauthoryear{Andr\'{e}, Ward-Thompson, \& Motte}
{Andr{\' e} et al.}{1996}]{1996A&A...314..625A} 
Andr\'{e} P., Ward-Thompson D., Motte F., 1996, A\&A, 314, 625 

\bibitem[\protect\citeauthoryear{Andr{\' e}, Ward-Thompson, \& Barsony}
{Andr{\' e} et al.}{2000}]{2000prpl.conf...59A} 
Andr{\' e} P., Ward-Thompson D., Barsony M., 2000, in Mannings V., Boss A., 
Russell S. S., eds, 
Protostars and Planets IV. Univ. Arizona Press, p. 59

\bibitem[\protect\citeauthoryear{Archibald et al.}{2002}]{2002MNRAS.336....1A}
Archibald E.~N., et al., 2002, MNRAS, 336,1 

\bibitem[\protect\citeauthoryear{Bacmann et al.}{2000}]{2000A&A...361..555B} 
Bacmann A., Andr{\' e} P., Puget J.-L., Abergel A., Bontemps S., Ward-Thompson
D., 2000, A\&A, 361, 555 

\bibitem[\protect\citeauthoryear{Beckwith et al.}{1990}]{1990AJ.....99..924B} 
Beckwith S.~V.~W., Sargent A.~I., Chini R.~S., Guesten R., 1990, AJ, 99, 924 

\bibitem[\protect\citeauthoryear{Bertout, Robichon, \& Arenou}{Bertout et al.}
{1999}]{1999A&A...352..574B} 
Bertout C., Robichon N., Arenou F., 1999, A\&A, 352, 574 

\bibitem[\protect\citeauthoryear{Bonnor}{1956}]{1956MNRAS.116..351B} 
Bonnor W.~B., 1956, MNRAS, 116, 351 

\bibitem[\protect\citeauthoryear{Bontemps et al.}{2001}]{2001A&A...372..173B}
Bontemps S., et al., 2001, A\&A, 372, 173 

\bibitem[\protect\citeauthoryear{Casanova et al.}{1995}]{1995ApJ...439..752C}
Casanova S., Montmerle T., Feigelson E.~D., Andre P., 1995, ApJ, 439, 752 

\bibitem[\protect\citeauthoryear{Castets et al.}{2001}]{2001A&A...375...40C}
Castets A., Ceccarelli C., Loinard L., Caux E., Lefloch B., 2001, A\&A, 375, 
40 

\bibitem[\protect\citeauthoryear{Ceccarelli et al.}{1998}]{1998A&A...331..372C}
Ceccarelli C., et al., 1998, A\&A, 331, 372

\bibitem[\protect\citeauthoryear{Correia, Griffin, \& Saraceno}{Correia et al.}
{2004}]{2004A&A...418..607C} 
Correia J.~C., Griffin M., Saraceno P., 2004, A\&A, 418, 607 

\bibitem[\protect\citeauthoryear{Ebert}{1955}]{1955ZA.....37..217E} 
Ebert R., 1955, ZA, 37, 217 

\bibitem[\protect\citeauthoryear{Elmegreen \& Lada}{1977}]{1977ApJ...214..725E}
Elmegreen B.~G., Lada C.~J., 1977, ApJ, 214, 725 
 
\bibitem[\protect\citeauthoryear{Emerson}{1995}]{1995mfsr.conf..309E} 
Emerson D.~T., 1995, in ASP Conf. Ser. 75, Multifeed Systems for Radio 
Telescopes, ed. D. T. Emerson \& J. M. Payne (San Francisco: ASP), 309

\bibitem[\protect\citeauthoryear{Grasdalen, Strom, \& Strom}{1973}]
{1973ApJ...184L..53G}
Grasdalen G.~L., Strom K.~M., Strom S.~E., 1973, ApJ, 184, L53 

\bibitem[\protect\citeauthoryear{Greene et al.}{1994}]{1994ApJ...434..614G} 
Greene T.~P., Wilking B.~A., Andr\'{e} P., Young E.~T., Lada C.~J., 1994, 
ApJ, 434, 614 

\bibitem[\protect\citeauthoryear{Henning, Michel, \& Stognienko}
{Henning et al.}{1995}]{1995P&SS...43.1333H} 
Henning T., Michel B., Stognienko R., 1995, P\&SS, 43, 1333 

\bibitem[\protect\citeauthoryear{Hildebrand}{1983}]{1983QJRAS..24..267H} 
Hildebrand R.~H., 1983, QJRAS, 24, 267 

\bibitem[\protect\citeauthoryear{Jenness \& Lightfoot}{2000}]{SURF}
Jenness T., Lightfoot J.F., 2000, Starlink User Note 216, Starlink Project, 
CCLRC 

\bibitem[\protect\citeauthoryear{Johnstone et al.}{2000}]{2000ApJ...545..327J}
Johnstone D., Wilson C.~D., Moriarty-Schieven G., Joncas G., Smith G., 
Gregersen E., Fich M., 2000, ApJ, 545, 327

\bibitem[\protect\citeauthoryear{Johnstone, Di Francesco \& Kirk}
{Johnstone et al.}{2004}]{2004ApJ...611L..45J} 
Johnstone D., Di Francesco J., Kirk H., 2004, ApJ, 611, L45 

\bibitem[\protect\citeauthoryear{Kirk, Ward-Thompson, \& Andr{\'e}}
{Kirk et al.}{2005}]{2005MNRAS.360.1506K} 
Kirk J.~M., Ward-Thompson D., Andr{\'e} P., 2005, MNRAS, 360, 1506 
 
\bibitem[\protect\citeauthoryear{Lada}{1987}]{1987IAUS..115....1L} 
Lada C.~J., 1987, IAUS, 115, 1 

\bibitem[\protect\citeauthoryear{Lasker}{1994}]{1994IAUS..161..167L} 
Lasker B.~M., 1994, IAUS, 161, 167 

\bibitem[\protect\citeauthoryear{Loren}{1989}]{1989ApJ...338..902L} 
Loren R.~B., 1989, ApJ, 338, 902 

\bibitem[\protect\citeauthoryear{Loren \& Wootten}{1986}]{1986ApJ...306..142L}
Loren R.~B., Wootten A., 1986, ApJ, 306, 142 

\bibitem[\protect\citeauthoryear{Loren, Wootten, \& Wilking}{Loren et al.}{1990}]{1990ApJ...365..269L} 
Loren R.~B., Wootten A., Wilking B.~A., 1990, ApJ, 365, 269 

\bibitem[\protect\citeauthoryear{Lynds}{1962}]{1962ApJS....7....1L} 
Lynds B.~T., 1962, ApJS, 7, 1 

\bibitem[\protect\citeauthoryear{Martin et al.}{1998}]{1998MNRAS.300..733M} 
Martin E.~L., Montmerle T., Gregorio-Hetem J., Casanova S., 1998, MNRAS, 300, 
733 

\bibitem[\protect\citeauthoryear{McGlynn \& Scollick}{1994}]
{1994ASPC...61...34M} 
McGlynn T., Scollick K., 1994, ASPC, 61, 34 

\bibitem[\protect\citeauthoryear{Motte, Andr\'{e}, \& Neri}
{Motte et al.}{1998}]{1998A&A...336..150M} 
Motte F., Andr\'{e} P., Neri R., 1998, A\&A, 336, 150 

\bibitem[\protect\citeauthoryear{Mundy et al.}{1992}]{1992ApJ...385..306M} 
Mundy L.~G., Wootten A., Wilking B.~A., Blake G.~A., Sargent A.~I., 1992, 
ApJ, 385, 306 

\bibitem[\protect\citeauthoryear{Myers \& Benson}{1983}]{1983ApJ...266..309M}
Myers P.~C., Benson P.~J., 1983, ApJ, 266, 309 

\bibitem[\protect\citeauthoryear{Ossenkopf \& Henning}{1994}]
{1994A&A...291..943O} 
Ossenkopf V., Henning T., 1994, A\&A, 291, 943 

\bibitem[\protect\citeauthoryear{Pigulski}{1992}]{1992A&A...261..203P} 
Pigulski A., 1992, A\&A, 261, 203 

\bibitem[\protect\citeauthoryear{Preibisch \& Zinnecker}{2001}]
{2001ASPC..243..791P}
Preibisch T., Zinnecker H., 2001, ASPC, 243, 791 

\bibitem[\protect\citeauthoryear{Sandell}{1994}]{1994MNRAS.271...75S} 
Sandell G., 1994, MNRAS, 271, 75 

\bibitem[\protect\citeauthoryear{Stark et al.}{2004}]{2004ApJ...608..341S} 
Stark R., et al., 2004, ApJ, 608, 341 

\bibitem[\protect\citeauthoryear{Stine et al.}{1988}]{1988AJ.....96.1394S} 
Stine P.~C., Feigelson E.~D., Andre P., Montmerle T., 1988, AJ, 96, 1394 

\bibitem[\protect\citeauthoryear{Wakelam et al.}{2004}]{2004A&A...413..609W} 
Wakelam V., Castets A., Ceccarelli C., Lefloch B., Caux E., Pagani L., 2004, 
A\&A, 413, 609 

\bibitem[\protect\citeauthoryear{Walker, Carlstrom, \& Bieging}{1993}]
{1993ApJ...402..655W} 
Walker C.~K., Carlstrom J.~E., Bieging J.~H., 1993, ApJ, 402, 655 

\bibitem[\protect\citeauthoryear{Ward-Thompson et al.}{2006}]{PPV}
Ward-Thompson D., Andr\'e P., Crutcher R. M., Johnstone D.,
Onishi T., Wilson C., 2006, in: `Protostars and Planets V', in press

\bibitem[\protect\citeauthoryear{Ward-Thompson, Andr{\' e}, \& Kirk}
{Ward-Thompson et al.}{2002}]{2002MNRAS.329..257W} 
Ward-Thompson D., Andr{\' e} P., Kirk J.~M., 2002, MNRAS, 329, 257 

\bibitem[\protect\citeauthoryear{Ward-Thompson et al.}{1989}]
{1989MNRAS.241..119W} 
Ward-Thompson D., Robson E.~I., Whittet D.~C.~B., Gordon M.~A., 
Walther D.~M., Duncan W.~D., 1989, MNRAS, 241, 119 

\bibitem[\protect\citeauthoryear{Ward-Thompson et al.}{1994}]
{1994MNRAS.268..276W} 
Ward-Thompson D., Scott P.~F., Hills R.~E., Andr\'e P., 1994, MNRAS, 268, 276

\bibitem[\protect\citeauthoryear{Wilking \& Lada}{1983}]{1983ApJ...274..698W}
Wilking B.~A., Lada C.~J., 1983, ApJ, 274, 698 

\bibitem[\protect\citeauthoryear{Wilking, Lada, \& Young}{Wilking et al.}
{1989}]{1989ApJ...340..823W} 
Wilking B.~A., Lada C.~J., Young E.~T., 1989, ApJ, 340, 823 

\bibitem[\protect\citeauthoryear{Wootten}{1989}]{1989ApJ...337..858W} 
Wootten A., 1989, ApJ, 337, 858 

\end{thebibliography}
\end{document}